\documentstyle[12pt]{article}
\def\ave#1{\left\langle #1\right\rangle}
\def\half{{\textstyle\frac{1}{2}}}

\font\rm=cmr10 scaled\magstep1
\def\re{\hbox{\rm Re}}
\def\im{\hbox{\rm Im}}
\def\per{\hbox{\rm per}}
\def\sper{\hbox{\rm sper}}

\newcommand{\ve}[1]{\hbox{\boldmath{$#1$}}}

\newcommand{\ma}[1]{\hbox{\boldmath{\rm #1}}}
\newcommand{\ti}[1]{\tilde{#1}}

\begin{document}
\begin{flushright}
Preprint CAMTP/96-2\\
December 14, 1995\\
revised: May 21, 1996
\end{flushright}
\begin{center}
\vspace{0.3in}
{\Large\bf  Exact statistics of complex zeros for Gaussian random
polynomials with real coefficients}\\ 
\vspace{0.4in}
\large
Toma\v z Prosen\footnote{Present address:
Physics Department, Faculty for Mathematics and Physics,
University of Ljubljana, Jadranska 19, 61000 Ljubljana,
Slovenia}\\
\normalsize
\vspace{0.3in}
Unit\' e mixte de service de l'institut Henri Poincar\' e, CNRS --
universit\' e Pierre et Marie Curie, Paris\\
\vspace{0.15in}
and\\
\vspace{0.15in}
Center for Applied Mathematics and Theoretical Physics,\\
University of Maribor, Krekova 2, SLO-62000 Maribor, Slovenia\\
\end{center}
\vspace{0.5in}

\noindent{\bf Abstract}
$k-$point correlations of {\em complex zeros} for Gaussian ensembles of 
{\em Random Polynomials} of order $N$ with {\em Real} Coefficients 
(GRPRC) are calculated exactly, following an approach of Hannay \cite{H95} 
for the case of Gaussian Random Polynomials with Complex Coefficients (GRPCC).
It is shown that in the thermodynamic limit $N\rightarrow\infty$ of
Gaussian random holomorphic functions all the statistics converge to the their
GRPCC counterparts as one moves off the real axis, while close to the real 
axis the two cases are essentially different. Special emphasis is given to 1 
and 2 point correlation functions in various regimes.
\vspace{0.7in}

\noindent
The problem of statistics of zeros of random polynomials of order $N$, and
of random holomorphic functions as $N\rightarrow\infty$ in general, arises in
various contexts in quantum chaos \cite{BBL1,BBL2}. The motivation for this 
work was the problem of statistics of zeros of coherent state (Husimi) or 
Bargmann \cite{B61} representation of eigenstates of chaotic 
systems \cite{LV93,P95}. It has been conjectured \cite{LV93} that zeros of
Bargmann or Husimi representation of an eigenfunction of 1-dim 
classically chaotic 
system should be uniformly and randomly scattered over the classically chaotic 
region of phase space. Bargmann representation of an eigenstate is an entire 
analytic function in a complex phase space variable $z=q + ip$, sometimes it is
even a polynomial of a finite order, like for example in the case of spin
systems where the phase space manifold is a sphere parametrized by 
$(\theta,\phi)$ and $z=\cot(\theta/2)\exp(i\phi)$ is a stereographic projection.
The coefficients of a power series of such entire functions or polynomials
are just the coefficients of an expansion of the chaotic eigenstate in a
complete set of (say harmonic) wavefunctions. Applying the random matrix
theory one argues that these coefficients should be uncorrelated 
(real/complex in the presence/absence of anti-unitary symmetry)  
pseudorandom Gaussian variables. Thus one can introduce the statistical
ensembles of random polynomials of order $N$ (or random analytic
functions in the limit $N\rightarrow\infty$) and argue that statistical
properties of their zeros can be used as a model to describe statistical
properties of zeros of a Bargman representation of chaotic eigenstates of
real systems.
\\\\
Recently, Hannay \cite{H95} has calculated general $k$-point correlation
functions of zeros of a
random spin state in a coherent state representation which is described by the
random polynomial with uncorrelated complex Gaussian coefficients,
and solved the problem of statistics of zeros of GRPCC --- Gaussian random 
polynomials with {\em complex} coefficients in general. It has been demonstrated
numerically \cite{LS95,P95up} that his results on GRPCC provide a universal 
description of the statistics of zeros of Bargmann or Husimi representation of 
chaotic eigenstates for systems without an anti-unitary symmetry.
Here we adopt this 
approach and solve the general problem of statistics of zeros $z_k$ of GRPRC
--- random polynomials $f(z)$ of order $N$
\begin{equation}
f(z) = \sum_{n=0}^N a_n z^n = a_N\prod_{j=1}^N (z-z_j)
\label{eq:pol}
\end{equation}
with {\em real} (Gaussian) coefficients $a_n$. We argue that the obtained 
results may be used to describe statistics of zeros of eigenstates of 1-dim.
(and quantum Poincar\' e sections \cite{P95} and other 
reductions \cite{V95,S95} 
of 2-dim.) chaotic systems in Bargmann representation with {\em time reversal 
invariance} (or any other anti-unitary symmetry
\footnote{For a general anti-unitary symmetry, the coefficients of the
random polynomials (\ref{eq:pol}) are of the form $a_n = r_n e^{i\theta_n}$
where $r_n$ are real Gaussian random variables and $\theta_n$ are fixed
(nonrandom) phases (which determine the symmetry curve in complex $z$-plane,
such as in fig.6 of ref. \cite{BBL2}). 
Then one may use the same general approach 
described below, eqs. (\ref{eq:var}-\ref{eq:sp}).}) to the same extent 
as Gaussian orthogonal ensembles of random matrices can be used to
describe the Hamiltonian and the typical observables.
\\\\
In the literature one may find several results on the distribution of
complex zeros of random polynomials with either complex \cite{A66} or real
\cite{SV95} Gaussian coefficients (see also \cite{BBL2} and references 
therein). The formula for 1-point 
function given below (\ref{eq:1pf}) (in the special case where the variances
of all coefficients are equal) is equivalent to the theorem 1.1 of Shepp and 
Vanderbei \cite{SV95}.
\\\\
Take a $k-$tuple of complex numbers $\ve{z}=({z_1,\ldots,z_k})$.
Since $a_n$ are real Gaussian random variables (which in general need not be 
uncorrelated!), their real linear combinations 
\begin{equation}
f^r_j =\re f(z_j),\quad
f^i_j =\im f(z_j),\quad
f^{\prime r}_j = \re\frac{d}{dz}f(z_j),\quad
f^{\prime i}_j = \im\frac{d}{dz}f(z_j),\quad j=1,\ldots,k
\label{eq:var}
\end{equation}
are also real Gaussian random variables with a joint distribution
\begin{equation}
P(\ve{f}^r,\ve{f}^i,\ve{f}^{\prime r},\ve{f}^{\prime i}) = 
(\det{2\pi\ti{\ma{M}}})^{-1/2}\exp\left(-\half
(\ve{f}^r,\ve{f}^i,\ve{f}^{\prime r},\ve{f}^{\prime i})\cdot
\ti{\ma{M}}^{-1}(\ve{f}^r,\ve{f}^i,\ve{f}^{\prime r},\ve{f}^{\prime i})\right).
\end{equation}
$\ti{\ma{M}}$ is a $4k\times 4k$ real symmetric positive correlation matrix
\begin{equation}
\ti{\ma{M}} = \pmatrix{
\ave{f^r_j f^r_l}&\ave{f^r_j f^i_l}
   &\ave{f^r_j f^{\prime r}_l}&\ave{f^r_j f^{\prime i}_l}\cr
\ave{f^i_j f^r_l}&\ave{f^i_j f^i_l}
   &\ave{f^i_j f^{\prime r}_l}&\ave{f^i_j f^{\prime i}_l}\cr
\ave{f^{\prime r}_j f^r_l}&\ave{f^{\prime r}_j f^i_l}
   &\ave{f^{\prime r}_j f^{\prime r}_l}&\ave{f^{\prime r}_j f^{\prime i}_l}\cr
\ave{f^{\prime i}_j f^r_l}&\ave{f^{\prime i}_j f^i_l}
   &\ave{f^{\prime i}_j f^{\prime r}_l}&\ave{f^{\prime i}_j f^{\prime i}_l}\cr}
= 
\pmatrix{\ti{\ma{A}}&\ti{\ma{B}}\cr\ti{\ma{B}}^T&\ti{\ma{C}}\cr} 
\end{equation}
where $\ave{}$ denotes the Gaussian ensemble averages which can be calculated
using (\ref{eq:pol},\ref{eq:var}) in terms of input data $\ave{a_n a_m}$.
One can write the $k-$point correlation function $\rho_k(\ve{z})$
in the following form
\begin{equation}
\rho_k(\ve{z}) = 
\int P(\ve{0},\ve{0},\ve{f}^{\prime r},\ve{f}^{\prime i}) 
\prod_{j=1}^k 
[(f^{\prime r}_j)^2 + (f^{\prime i}_j)^2]df^{\prime r}_j df^{\prime i}_j 
\end{equation}
where the factors $(f^{\prime r}_j)^2 + (f^{\prime i}_j)^2$ are just the
Jacobians of transformations from the pairs of real variables
$(f^r_j,f^i_j)$ to complex variables --- zeros $z_j$.
The integral can be written in terms of derivatives of a generating function
$Z_k(\ve{u},\ve{v})$
\begin{equation}
\rho_k(\ve{z}) = (-1)^k\prod_{j=1}^k (\partial^2_{u_j}+\partial^2_{v_j})
Z_k(\ve{u},\ve{v})\vert_{\ve{u}=\ve{v}=0}
\label{eq:rhog1}
\end{equation}
which is an ordinary Gaussian integral and can be explicitly 
calculated
\begin{eqnarray}
Z_k(\ve{u},\ve{v}) &=& (\det 2\pi\ti{\ma{M}})^{-1/2}
\int\exp\left(-\half(\ve{f}^{\prime r},\ve{f}^{\prime i})\cdot
\ti{\ma{L}}(\ve{f}^{\prime r},\ve{f}^{\prime i}) 
+ i\ve{f}^{\prime r}\cdot\ve{u} + i\ve{f}^{\prime i}\cdot\ve{v} \right)
\prod_{j=1}^k df^{\prime r}_j df^{\prime i}_j
\nonumber \\
&=&(\det 2\pi\ti{\ma{A}})^{-1/2}\exp\left(-\half (\ve{u},\ve{v})\cdot
\ti{\ma{L}}(\ve{u},\ve{v})\right)
\label{eq:rhog2}
\end{eqnarray}
where $\ti{\ma{L}} = \ti{\ma{C}} - \ti{\ma{B}}^T \ti{\ma{A}}^{-1}\ti{\ma{B}}$ 
is a lower right block of $\ti{\ma{M}}^{-1}$ and we have used an identity 
\cite{H95} $\det\ti{\ma{L}}/\det\ti{\ma{M}} = 1/\det{\ti{\ma{A}}}$.
At this point it is convenient to switch on the equivalent complex variables
$\ve{f}=\ve{f}^r + i \ve{f}^i,\,
\ve{f}^\prime=\ve{f}^{\prime r} + i \ve{f}^{\prime i},\,
\ve{w}=\ve{u}+i\ve{v}$ and their complex conjugates.
Then one can write eq. (\ref{eq:rhog1},\ref{eq:rhog2}) as
\begin{eqnarray}
\rho_k(\ve{z}) &=& \frac{(-1)^k 2^k}{(\det 2\pi\ma{A})^{1/2}}
\prod\limits_{j=1}^k \partial_{w_j}\partial_{w^*_j}
\exp\left(-\half (\ve{w}^*,\ve{w})\cdot\ma{L}(\ve{w},\ve{w}^*)\right)
\vert_{\ve{w}=0}\nonumber\\
&=&(\det 2\pi\ma{A})^{-1/2} 
\prod\limits_{j=1}^k \partial_{w_j}\partial_{w^*_j}
\left((\ve{w}^*,\ve{w})\cdot\ma{L}(\ve{w},\ve{w}^*)\right)^k
\vert_{\ve{w}=0}
\label{eq:rrr}
\end{eqnarray}
where all the $2k\times 2k$ real matrices should be transformed by the rule
$$\ma{X}=\ma{U}^\dagger\ti{\ma{X}}\ma{U},\quad
\ma{U}=\half\pmatrix{\ma{1}&\ma{1}\cr i\ma{1}&-i\ma{1}\cr}$$
giving $\ma{L} = \ma{C} - \ma{B}^\dagger\ma{A}^{-1}\ma{B}$ with
\begin{eqnarray}
\ma{A} &=& \pmatrix{\ave{f_j f^*_k} & \ave{f_j f_k}\cr
                  \ave{f^*_j f^*_k} & \ave{f^*_j f_k}\cr} = \ma{A}^\dagger,\\
\ma{B} &=& \pmatrix{\ave{f_j f^{\prime *}_k} & \ave{f_j f^\prime_k}\cr
                  \ave{f^*_j f^{\prime *}_k} & \ave{f^*_j f^\prime_k}\cr},\\
\ma{C} &=& \pmatrix{\ave{f^\prime_j f^{\prime *}_k} 
                  & \ave{f^\prime_j f^\prime_k}\cr
                  \ave{f^{\prime *}_j f^{\prime *}_k} 
                  & \ave{f^{\prime *}_j f^\prime_k}\cr} = \ma{C}^\dagger.
\end{eqnarray}
Applying a little combinatorics on eq. (\ref{eq:rrr}) we finally obtain the 
general result
\begin{equation}
\rho_k(\ve{z}) = 
\frac{\sper\left(\ma{C} - \ma{B}^\dagger\ma{A}^{-1}\ma{B}\right)}
{\sqrt{\det 2\pi \ma{A}}}
\label{eq:genres}
\end{equation}
where we introduce the {\em semi-permanent} of a $2k\times 2k$ matrix
\begin{equation}
\sper\ma{L} =
\sum_{j_1 < \ldots < j_k \atop l_1 < \ldots < l_k}^{j_m \neq l_n}
\sum\limits_{p\in S_k} \prod\limits_{r=1}^k L_{j_r + k,l_{p(r)}} 
\label{eq:sp}
\end{equation}
The first sum runs over $(2k)!/(k!)^2$ ordered combinations of 
$k$ out of $2k$ indices $j_m$ and their complements $l_n$ while the second
sum runs over $k!$ permutations $p$ of the symmetric group $S_k$. The sum of
indices $j_r + k$ should be taken modulo $2k$.
\\\\
So far we have not assumed anything about the correlations between the
coefficients $a_n$ expect the Gaussian nature of the joint distribution of 
coefficients $a_n$.
Now we shall asume that Gaussian coefficients $a_n$ are uncorrelated and 
define the polynomial $g(s)$ with positive coefficients $b_n$ --- the 
variances of $a_n$
\begin{eqnarray}
\ave{a_n a_m} &=& b_n\delta_{nm},\quad b_n > 0,\\
g(s) &=& \sum_{n=0}^N b_n s^n.
\end{eqnarray}
The matrices $\ma{A}$,$\ma{B}$, and $\ma{C}$ can be easily expressed solely
in terms of a polynomial $g$ and its derivatives $g^\prime,g^{\prime\prime}$
\begin{eqnarray}
A_{j l}(\ve{z}) &=& g(z_j z^*_l),\\
B_{j l}(\ve{z}) &=& 
\partial_{z^{*}_l}A_{j l}(\ve{z}) = z_j g^{\prime}(z_j z^{*}_l),\\
C_{j l}(\ve{z}) &=&
\partial_{z_j}\partial_{z^{*}_l}A_{j l}(\ve{z}) =
g^{\prime}(z_j z^{*}_l) + z_j z^{*}_l g^{\prime\prime}(z_j z^{*}_l)
\end{eqnarray}
where we let indices $j,l$ to run from $1$ through $2k$ and put 
$z_{k+j}:= z^*_{j}$. Note that the time-reversal symmetry --- the symmetry of 
zeros with respect to the reflection
over the real axis is present also in the $k-$point correlation functions,
namely
$$\rho_k(z_1,\ldots,z_j,\ldots,z_k) = \rho_k(z_1,\ldots,z_j^*,\ldots,z_k).$$
Without loss of generality one may assume that all points $z_j$ lie on the
upper complex halfplane, $\im z_j > 0$. Otherwise one gets long range 
correlations in cases where one of the points $z_j$ comes close to the mirror 
image of one of the other points $z^*_l$.

In general, only the 1-point function $\rho_1(z)$ ---- the density of zeros 
is simple enough to be written out
\begin{equation}
\rho_1(z) = \frac{g^{\prime}_0 + |z|^2 g^{\prime\prime}_0}
{\pi(g^2_0 - g_+ g_-)^{1/2}}
+ \frac{(z^2 g_- g^{\prime}_+ + {z^{*}}^2 g_+ g^{\prime}_-)
g^{\prime}_0 - |z|^2 (g^{\prime}_+ g^{\prime}_- + {g^{\prime}_0}^2) g_0}
{\pi(g^2_0 - g_+ g_-)^{3/2}},
\label{eq:1pf}
\end{equation}
where 
$g_0 \equiv g(|z|^2),\; g_+ \equiv g(z^2),\; g_- \equiv g({z^*}^2).$
Writing $z=x+iy$ and carefully expanding for small $y$ one finds
\begin{eqnarray}
\rho_1(z) &=& h(x^2)|y| + {\cal O}(y^3), \quad y \neq 0 \label{eq:1pfe}\\
h(s) &=& (2\pi)^{-1}
(g g^\prime - s {g^\prime}^2 + s g g^{\prime\prime})^{-3/2}
(2g_{012} + 2(2g_{013}-g_{112}-g_{022})s \nonumber\\
&& + (3g_{122}-4g_{113}+g_{014})s^2 + 
(g_{024}-g_{114}-g_{033}+2g_{123}-g_{222})s^3)\nonumber
\end{eqnarray}
where $g\equiv g(s),\;g_{nml}\equiv g^{(n)}(s)g^{(m)}(s)g^{(l)}(s)$.
So quite generally, the density of zeros decreases linearly as we approach the
real axis. To evaluate the density of zeros on a real axis $y=0$ one should
use a different approach described in \cite{BBL2}.
In another asymptotical regime $|z|\rightarrow\infty$, only the highest power
terms of $g$ contribute, and one finds
\begin{equation}
\rho_1(z) = \frac{2 b_{N-2}}{\sqrt{b_N b_{N-1}}} \frac{\im z}{|z|^6}
\left(1 + {\cal O}\left(\frac{1}{|z|^2}\right)\right).
\end{equation}
So, the density of zeros vanishes asymptotically
since the total number of zeros $N$ is finite.
\\\\
Now we shall study the thermodynamic limit $N\rightarrow\infty$.
It is convenient to study {\em random holomorphic functions} which provide 
a uniform distribution of zeros in the complex plane. A unique choice 
(up to rescaling $s\rightarrow\lambda s$) is $b_n = 1/n!$ giving 
\begin{equation}
g(s) = \exp(s).
\end{equation}
Such random holomorphic functions naturally arise when one studies 
Bargmann representation of 1-dim chaotic eigenstates in the usual 
$(p,q)\in \Re^2$ phase space.
We argue that any other choice will only affect the density of zeros
$\rho_1(z)$ while properly rescaled local statistics should be independent
on the choice of $g(s)$ provided that variances of coefficients
$b_n$ {\em depend smoothly} on $n$.

Away enough from the real axis $\im z_j \gg 1$ one may neglect the
offdiagonal $k\times k$ blocks of matrices $\ma{A},\ma{B},\ma{C}$
since the ratios of the corresponding matrix elements become
exponentially small $|\exp(z_j z^*_l)/\exp(z_j z_l)| = 
\exp(-2 \im z_j \im z_l)$.
Then using straightforward results
\begin{eqnarray}
2^{-k}\sper\pmatrix{\ma{L}_{11} &\ma{0}\cr\ma{0} &\ma{L}^T_{11}\cr} &=&
\per\ma{L}_{11} := \sum\limits_{p\in S_k}\prod_{j=1}^k L_{j,p(j)}\\
\det\pmatrix{\ma{A}_{11} &\ma{0}\cr\ma{0} &\ma{A}^T_{11}\cr} &=&
\left(\det\ma{A}_{11}\right)^2,
\end{eqnarray}
where $()_{11}$ denotes the upper-left $k\times k$ block of a $2k\times 2k$
matrix, one arrives to the result which is equivalent to the statistics of 
zeros of GRPCC \cite{H95} 
\begin{equation}
\rho_k(\ve{z})\rightarrow
\rho^{\hbox{\rm GRPCC}}_k(\ve{z}) = 
\frac{\per (\ma{C}_{11} - \ma{B}^\dagger_{11}\ma{A}^{-1}_{11}\ma{B}_{11})}
{\det \pi\ma{A}_{11}},\quad\hbox{\rm as}\quad \im z_j\rightarrow\infty.
\label{eq:Hannay}
\end{equation}
\\\\
To conclude we give some explicit results about 1 and 2 point functions.
The density of zeros which is shown in figure 1 reads
\begin{equation}
\rho_1(x+iy) = \frac{1 - (4y^2+1)\exp(-4 y^2)}{\pi (1 - \exp(-4 y^2))^{3/2}},
\label{eq:dens}
\end{equation}
which is a constant $1/\pi$ provided that we are away 
enough from the real axis. The excess of zeros due to the presence of real 
axis $\int_{-\infty}^\infty (1/\pi - \rho_1(x+iy))dy = 1/\pi$ is on the other 
hand just the linear density of real zeros on the real axis!

The 2-point correlation function $\rho_2(z_1,z_2)$ is already too lengthy to
be written out in general.
The behaviour of a normalized 2-point correlation function
$\rho_2(z_1,z_2)/\rho_1(z_1)/\rho_1(z_2)$ as we approach the real axis,
is shown in figure 2, while far away $\im z_1,\im z_2\gg 1$ 
it becomes isotropic and the result for GRPCC \cite{H95} applies
\begin{eqnarray}
\rho_2(z_1,z_2) &\rightarrow& \varphi(|z_1-z_2|^2),\nonumber\\
\varphi(s) &=& \frac{\exp(-2s)(\exp(s)-1-s)^2+\exp(-s)(\exp(-s)-1+s)^2}
{\pi^2 (1-\exp(-s))^3}
\label{eq:r2}
\end{eqnarray}
In the asymptotic regime $\im z_j \gg 1$ one can calculate also the 
{\em number variance}
$\Sigma_2(r)$: the variance of the number of zeros ${\cal N}(r)$ inside
a circle of radius $r$
\begin{equation}
\Sigma_2(r) = \ave{{\cal N}^2(r)} - \ave{{\cal N}(r)}^2.
\end{equation}
It can be expressed in terms of a four-fold integral (over $z_1,z_2$) of a
2-point correlation, which can be reduced using eq. (\ref{eq:r2}) to a 
single integral
\begin{equation}
\Sigma_2(r) = r^2(1-r^2) + 
8\pi r^4\int_0^1 (\arccos\sqrt{t} - \sqrt{t(1-t)})\varphi(4 r^2 t)dt.
\label{eq:nv1}
\end{equation}
The number variance $\Sigma_2(r)$ 
starts as ``Poissonian'' $\ave{{\cal N}(r)}=r^2$ for small $r$ 
whereas for larger $r$ it has a linear asymptotics (see figure 3)
\begin{equation}
\Sigma_2(r) = \sigma r + {\cal O}(1/r) \approx \sigma\sqrt{\ave{{\cal N}(r)}},
\quad
\sigma = \frac{4}{\pi}\int\limits_0^\infty s^2(1 - \pi^2\varphi(s^2))ds \approx
0.36847.
\label{eq:nv2}
\end{equation}
Note that this formula (\ref{eq:nv1},\ref{eq:nv2}) is valid also for
GRPCC in general.
\\\\
In the present paper the statistics of zeros of Gaussian random polynomials with
real coefficients have been solved analytically (\ref{eq:genres}) 
following an approach of
Hannay for the case of complex coefficients. Several important special cases
have been considered in detail: (i) the case of mutually uncorrelated 
coefficients, which corresponds to the Bargmann representation of chaotic 
eigenstates in random matrix regime, has been studied and it has been shown 
that all $k-$point correlation functions converge to those of random polynomials
with complex coefficients derived by Hannay as all points $z_j,j=1\ldots k$ 
move away from the real axis $Im z_j \gg 1$ (\ref{eq:Hannay}),
(ii) 1-point function -- the density of zeros have been written out in general
(eqs. (\ref{eq:1pf},\ref{eq:dens}) and fig. 1) and linear decrease of density 
towards the symmetry
line -- real axis has been found (\ref{eq:1pfe}), (ii) 2-point function close 
to the real axis have been explored numerically (fig. 2) while simple analytic 
formula (\ref{eq:r2}), which holds far away from the
real axis (and holds generally in the case of complex coefficients), 
have been used to 
derive a simple expression for the number variance of zeros inside a circle of 
a given radius (eqs. (\ref{eq:nv1},\ref{eq:nv2}) and fig. 3).
\\\\
Discussions with P.Leboeuf, J.H.Hannay, M.Saraceno and K.{\.Z}yczkowski as 
well as the hospitality of Institut Henri Poincar\' e (Paris) are gratefully
acknowledged. This work has been financially supported by the C.I.E.S (France)
and the Ministry of Science and Technology of the Republic of Slovenia.

\newpage
\section*{Figure captions}

\bigskip\noindent {\bf Fig.1} We show the density of zeros $\rho_1(x+iy)$ in 
the thermodynamic limit $N\rightarrow\infty$ given by eq. (\ref{eq:dens}) as
a function of the distance from the real axis.

\bigskip\noindent {\bf Fig.2} 
The normalized 2-point correlation function 
$\rho_2(x_1+iy,x_2+iy)/\rho_1(x_1+iy)/\rho_1(x_2+iy)$ 
in the limit $N\rightarrow\infty$ between two points, $x_1+iy$ and $x_2+iy$,
which have the same distance from the real axis $y$ is shown as a function
of $|x_2-x_1|$ for different values of $y=0.1,0.3,0.5,0.7,0.9,1.1,1.3,1.5.$
Note that all curves go to zero as $\propto y^2$ and that for $y\ge 1.5$
the 2 point correlation function has practically converged to the isotropic 
asymptotic one.

\bigskip\noindent {\bf Fig.3} The number variance $\Sigma_2(r)$ in the
asymptotical regime $N\rightarrow\infty,\im z\gg 1$ is shown as a function
of radius $r$ (\ref{eq:nv1}). 
\end{document}